\documentclass{ws-procs9x6}

\begin{document}

\title{DEEPLY VIRTUAL PSEUDOSCALAR MESON PRODUCTION WITH CLAS}
\author{V. Kubarovsky$^{1,2,\dag}$,  P. Stoler$^2$ and I. Bedlinsky$^3$} 

\address{\it $^1$Jefferson Lab, Newport News, USA,\\
$^2$Rensselaer Polytechnic  Institute, Troy, USA, \\
$^3$Institute for Theoretical and Experimental  Physics, Moscow, Russia\\
$^\dag$Email: vpk@jlab.org}
\begin{center}
\end{center}

\bodymatter

\subsection*{Introduction.}

Deeply virtual exclusive reactions offer a unique opportunity to study the 
structure of the nucleon at the parton level as one varies both the size of 
the probe, i.e. the photon virtuality $Q^2$, and the momentum transfer to the 
nucleon $t$. Such processes can reveal much more information about the 
structure of the nucleon than either inclusive electroproduction ($Q^2$ only) 
or elastic form factors ($t=- Q^2$ ). The characterization  of  deeply virtual
exclusive reactions in terms of their common nucleon structure is one of the major 
objectives of the Jefferson Lab 12 GeV upgrade.

There have been two  commonly used theoretical tools  which relate  exclusive 
reactions to the structure of the nucleon. At lower $Q^2$, where the probe is 
on the order of the size of
hadrons and the interactions are strong, Regge phenomena have
proved  effective. At high $Q^2$ the probe interacts
with individual quarks, and in  the limit $Q^2 \to \infty$ 
and $-t/Q^2\to 0$ the QCD factorization theorem~\cite{factorization}  
 unifies all exclusive reactions in terms of their
common nucleon structure encoded by generalized parton distributions (GPDs).
In this article we consider deeply virtual meson production (DVMP),
specifically the reaction $\gamma^* p \to p M$, where M is a meson 
($\pi,\eta,\rho,\omega, \phi, etc$).
In  the GPD approach, which is schematically shown in  
Fig.~1,
the ingredients involve  a hard interaction between a virtual photon and quark which
produces a meson whose internal structure is given by the distribution amplitude $\Phi(z)$,
and  the remaining nucleon whose  structure is represented  by GPDs. A caveat is
that the proof for factorization applies only to the case when the virtual photon has longitudinal 
polarization. In that case, in the limit $Q^2 \to \infty$\   the cross section 
scales as $\sigma_L \sim 1/Q^6$  and the ratio as $\sigma_T/\sigma_L\sim 1/Q^2$. 

\begin{figure}[b!]
\begin{center}
\begin{tabular}{c}
\mbox{\epsfig{figure=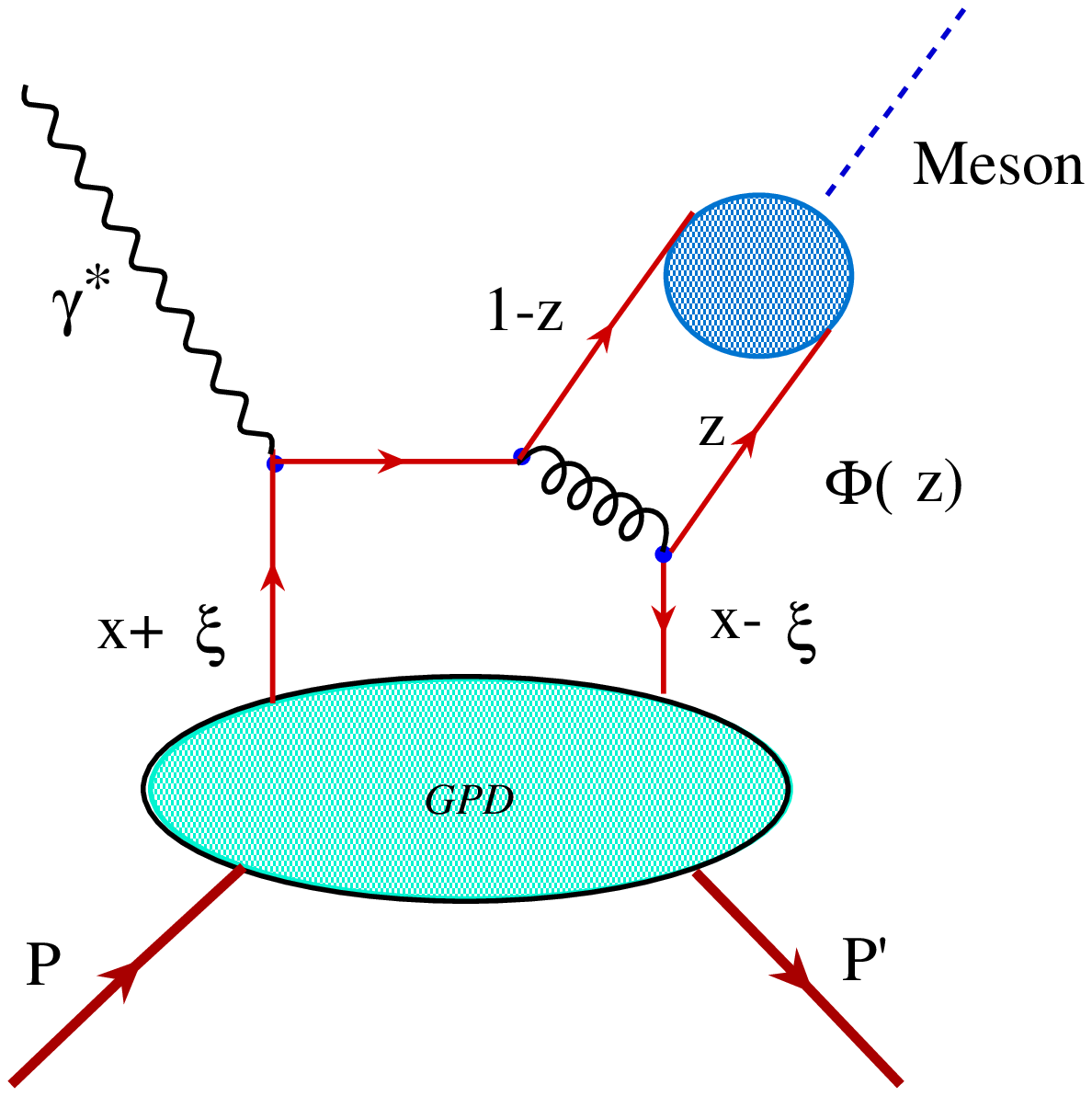,width=6cm,height=6cm}}\\
\end{tabular}
\end{center}
{\small{\bf Figure 1.}
Schematic illustration of the GPD approach to meson electroproduction.}\\
\label{gpd-meson}
\end{figure}


While most theoretical work  on the GPD approach has focused on the high $Q^2$ and low $|t|$ 
kinematic region, exclusive production of photons and mesons at large $|t|$
can also be described in terms the nucleon GPDs. Theory also predicts  
$\sigma_L$\  and $\sigma_T$  in the high $-t$ low $Q^2$ region.
For example for the pseudoscalar meson electroproduction we have:
\begin{equation}
{\frac{d\sigma}{dt}} \propto \left(R(t) \int dz \Phi(z)f(z,s,Q^2,t)\right )^2
\label{eq:gpd}
\end{equation}
$$R(t)\propto (e_u R^u(t) - e_d R^d(t)^d)    \ \ \ \ \    R^q(t)= \int{\frac{dx}{x}}
e^{\alpha t((1-x)/2x} \left[ \Delta q(x)-\Delta \bar q(x)\right]$$
\noindent where $\Phi(z)$ is the meson distribution amplitude, $f(z,s,Q^2,t)$
is parton  level amplitude and the $R(t)$s describe new form factors which are 
the $1/x$ moments of the GPDs. $\Delta q$ and   $\Delta \bar q$ are  the polarized 
parton and antiparton  distributions for $u,d$ and $s$ quarks. The constant in 
the exponent  $\alpha$\  is approximately 1 GeV$^{-2}$. The Fourier transforms 
with respect to $\overrightarrow{\Delta}_\perp$ ($\Delta^2=-t$) describe 
the correlation between the transverse spatial distribution   of 
quark impact and $x_B$\ in the proton.

Deeply virtual Compton scattering (DVCS) is the cleanest way of accessing GPDs. 
However, DVCS does not 
allow to discriminate 
the helicity dependent GPDs and it is difficult to 
perform a flavor separation.
In the case of pseudoscalar meson production the amplitude involves the axial 
vector-type GPDs 
$\tilde H$\ and $\tilde E$. These GPDs are closely related to the distribution
of quark spin in the proton, and $\tilde H$ reduces to the polarized quark/antiquark 
densities in the limit of zero  momentum transfer.
Vector and pseudoscalar meson production allows one to separate flavor and isolate 
the helicity-dependent GPDs. This is summarized  in Table~\ref{flavor}.

\begin{table}[h]
\begin{centering}
\begin{tabular}{|c|c|c|}
\hline
                                    & $\pi^+$ & $\Delta u - \Delta d$\\
$\tilde H$, $\tilde E$ & $\pi^0$ & $2\Delta u + \Delta d$\\
                                      & $\eta$   & $ 2\Delta u - \Delta d$\\
\hline      
                                     & $\rho^+$ & $ u -  d$\\
$H$, $E$                    & $\rho^0$ & $2 u + d$\\
                                      & $\omega$   & $ 2u -  d$\\
\hline      
\end{tabular}  
\caption{GPDs and quark flavor selectivity of pseudoscalar and vector meson electroproduction. }
\label{flavor}    
\end{centering}
\end{table}
 
The extraction of GPDs from electroproduction data is a challenging problem. A detailed 
understanding of the reaction mechanism is essential before one can compare with 
theoretical calculations. It is not yet clear at what values of $Q^2$ the application
of GPDs to meson electroproduction becomes valid.  However, detailed measurements 
of observables may test 
model-independent features of the reaction mechanism, such as $t$-slopes, 
flavor ratios, and generally by studying the variation of observables over a wide range of $Q^2$ and $t$.  
Even though current experiments are limited in $Q^2$ and $t$, it has been
argued~\cite{precocious} that  {\it precocious factorization} ratios of cross sections as a function of $x_B$
could be valid  at relatively lower $Q^2$  than for cross sections themselves.
For example, the ratio of  cross sections for  $\pi^0$ and $\eta$ electroproduction 
from a proton is related  to the helicity structure of the quark flavors as
\begin{equation}
\label{eq:pi0eta}
\pi^0/\eta = {\frac{1}{2}}\left[{\frac{2}{3}}\Delta u+{\frac{1}{3}}\Delta d\right ]^2/
{\frac{1}{6}}\left[{\frac{2}{3}}\Delta u-{\frac{1}{3}}\Delta d + {\frac{1}{3}}\Delta s\right ]^2 
\end{equation}
\noindent The results of a calculation of this ratio as a function of $x_B$ is 
shown in 
Fig.~2.
\begin{figure}[b!]
\begin{center}
\begin{tabular}{c}
\mbox{\epsfig{figure=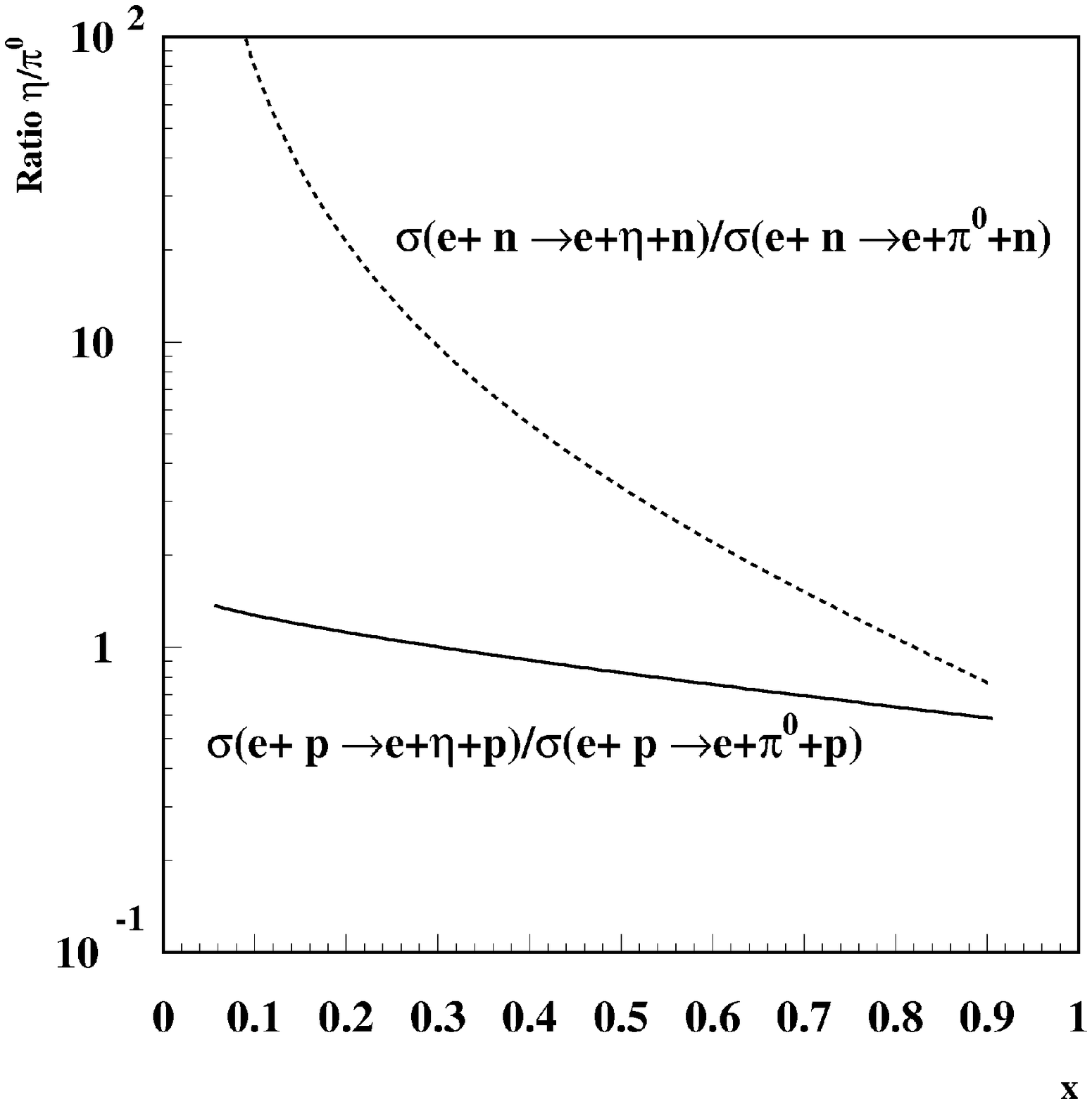,width=8cm,height=8cm}}\\
\end{tabular}
\end{center}
{\small{\bf Figure 2.} 
Predictions of the ratio of cross sections for $\pi^0$ to $\eta$ 
electroproduction from protons and neutrons~\cite{precocious} utilizing the 
concept of {\it precocious factorization.}}\\
\label{ratio}
\end{figure}



\subsection*{Recent CLAS measurements of $\pi^0$ and $\eta$ production.}
The beam spin asymmetries for  DVCS \cite{dvcs} and DVMP \cite{pi0} 
have recently been obtained  at Jefferson 
Lab with the
CLAS spectrometer, up to a $Q^2\sim 5$ GeV$^2$ 
(see also previous CLAS publications \cite{stepan} and \cite{chen}).  
This has been made 
possible by constructing a high quality  electromagnetic calorimeter 
consisting of  424 lead-tungsten crystals covering an angular range
from 4.5$^\circ$ to 15$^\circ$, which was positioned into the existing CLAS 
large acceptance detector. 
The pions and etas are identified through their 2$\gamma$ decays.
A photograph of the new detector and the   2$\gamma$ invariant mass distribution
is shown in 
Fig.~3.
One can see that the pions and etas are clearly
observed, even before all final data selection cuts are performed.
The kinematic coverage in the variables $Q^2, x_B, t$ and $W$ is shown in 
Fig.~4.

\begin{figure}[b!]
\begin{center}
\begin{tabular}{cc}
\mbox{\epsfig{figure=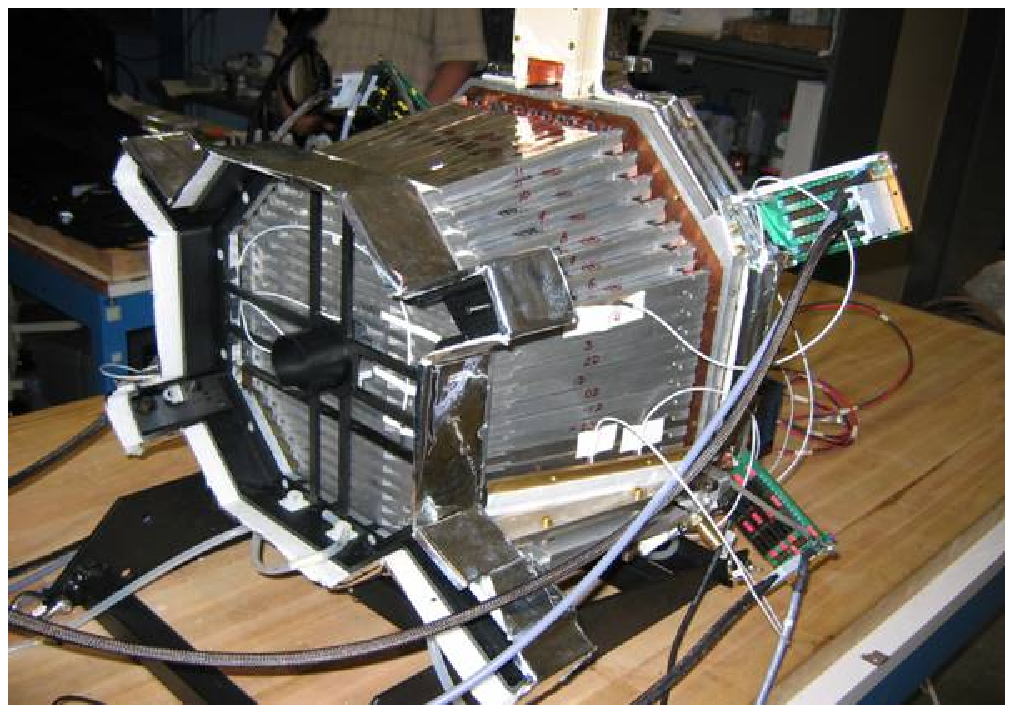,width=5cm,height=5cm}}&
\mbox{\epsfig{figure=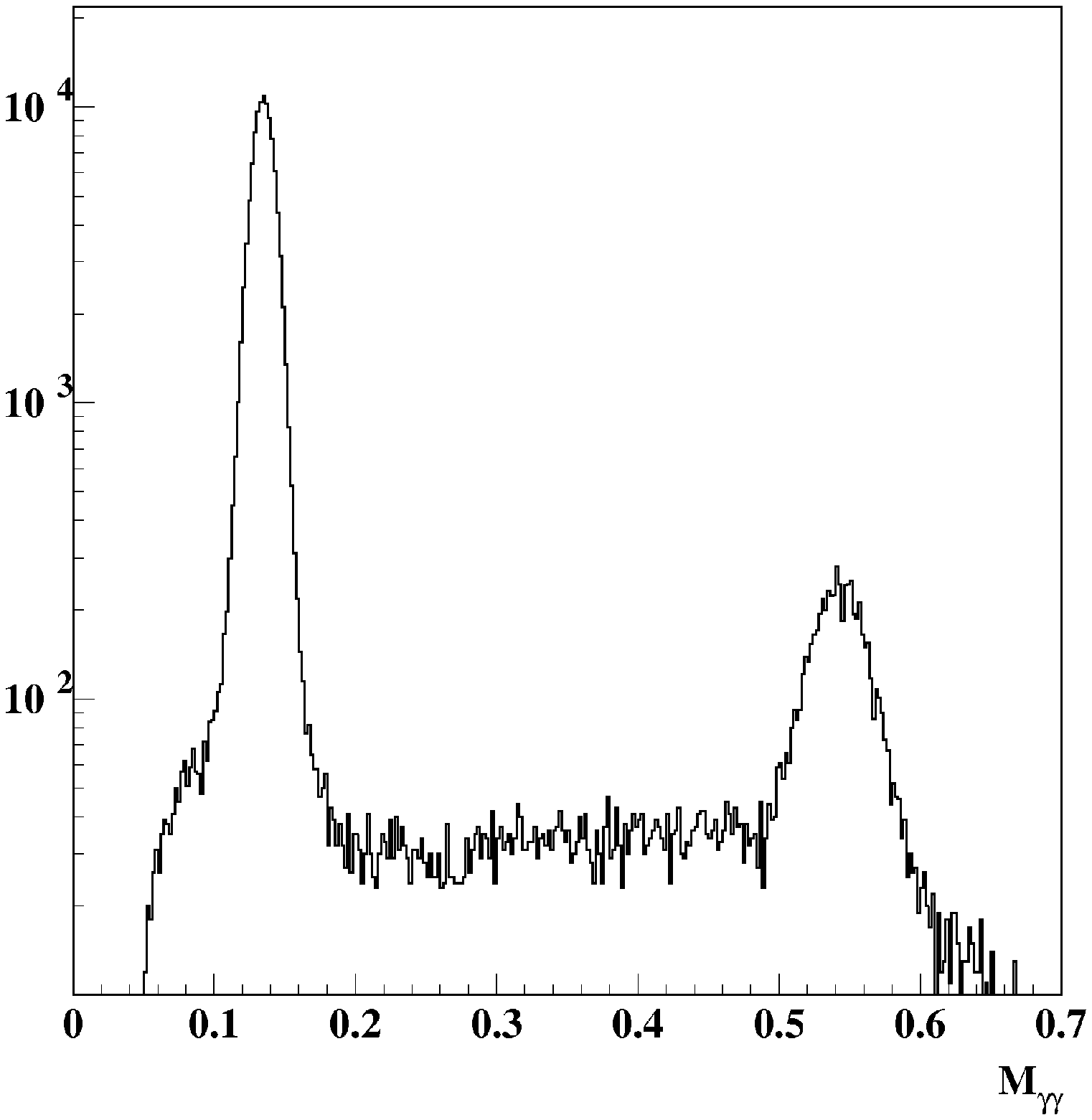,width=6cm,height=6cm}}\\
{\bf(a)}& {\bf(b)}
\end{tabular}
\end{center}
{\small{\bf Figure 3a.} 
Photograph of the new CLAS lead-tungsten electromagnetic calorimeter.}\\
{\small{\bf Figure 3b.} 
2$\gamma$ invariant mass spectrum in which the $\pi^0$'s and $\eta$'s are clearly
observed  (note log scale).}
\label{IC}
\end{figure}
\begin{figure}[b!]
\begin{center}
\begin{tabular}{ccc}
\mbox{\epsfig{figure=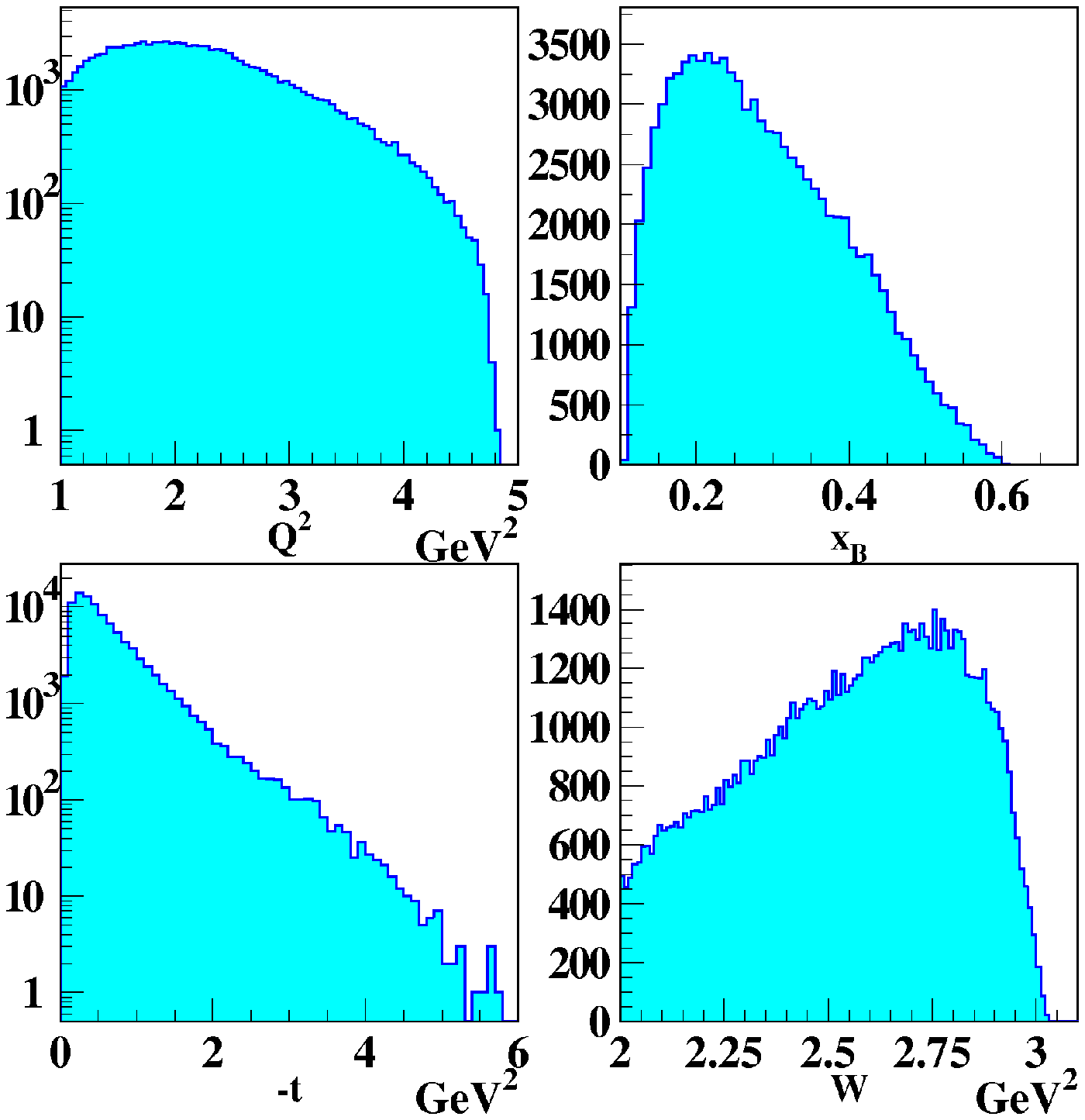,width=6cm,height=6cm}}\\
\end{tabular}
\end{center}
{\small{\bf Figure 4.} 
The kinematic coverage in $Q^2, t, x_B$ and $W$ for neutral pions  
of the CLAS DVMP experiment.
}
\label{coverage}
\end{figure}

 
The  virtual photon cross section can be written in well known notation as
\begin{equation}
\label{eq:sigma}
{ \frac{{d\sigma}}{{ d\Omega_\pi}  }}  = 
\sigma_T+\epsilon \sigma_L+ \epsilon \sigma_{TT} cos2\phi +  \sqrt{2\epsilon(1+\epsilon)/2}\sigma_{LT} cos\phi_\pi + h \sqrt{\epsilon(\epsilon-1)}  \sigma^\prime_{LT} sin\phi 
\end{equation}
\noindent where $\phi$ denotes the azimuthal angle between the hadronic and leptonic
scattering planes and $h$ is the  electron beam polarization. 

The large acceptance of CLAS enabled the data to be  grouped into 
intervals in 
$Q^2$, $t$\, $x_B$ and $\phi$.
For unpolarized electrons
($h=0$) the  separation of the $\phi$ dependence in moments of a $constant$, $cos\phi$,
and $cos2\phi$ allows us to obtain   $\sigma_T+\epsilon \sigma_L $, $ \sigma_{TT}$
and $\sigma_{LT}$.  An example of a $\phi$ distribution
for $t=0.3$ GeV$^2$ integrated over $Q^2$    are shown in 
Fig.~5.

\begin{figure}[b!]
\begin{center}
\begin{tabular}{ccc}
\mbox{\epsfig{figure=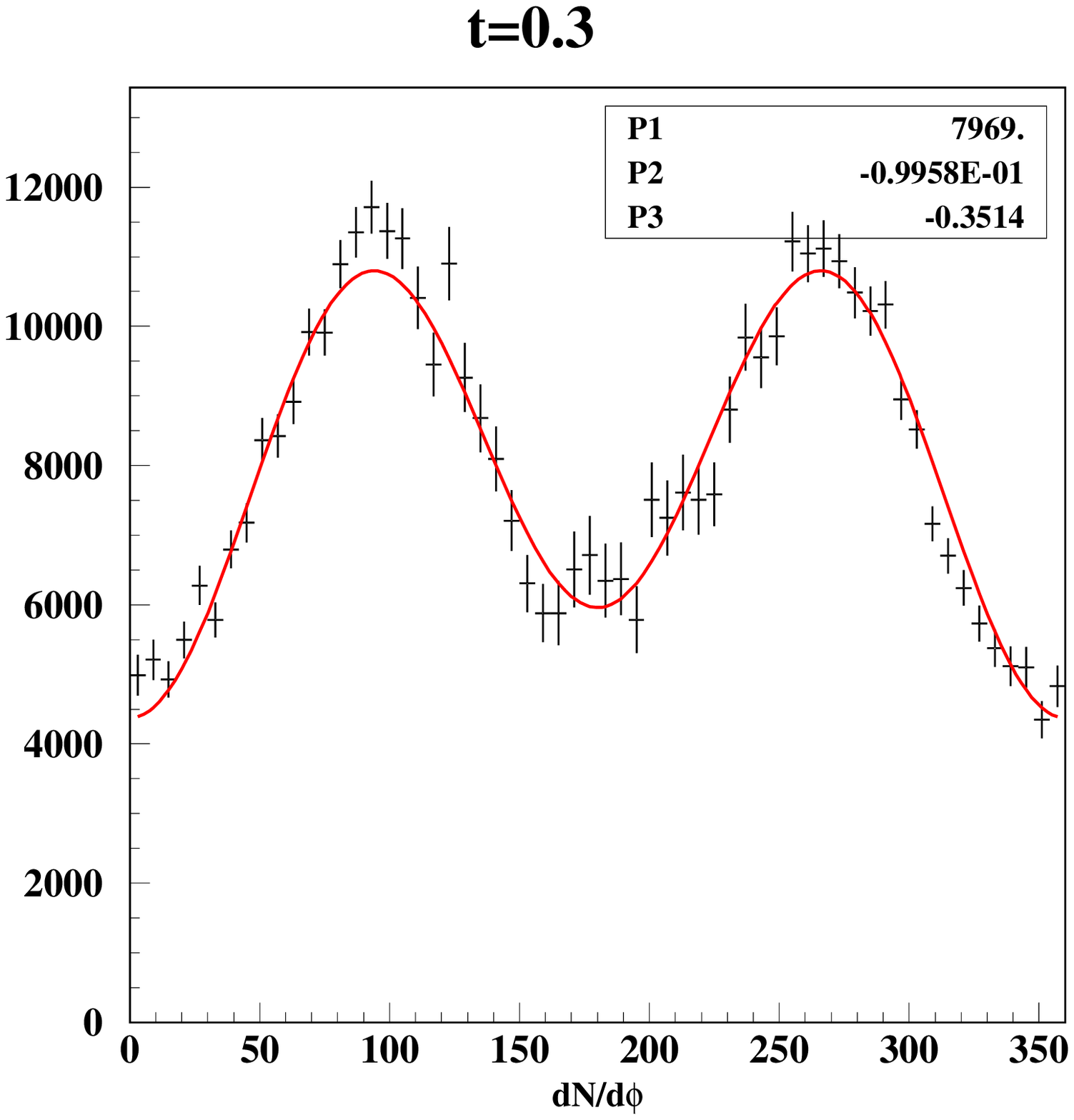,width=6cm,height=6cm}}\\
\end{tabular}
\end{center}

{\small{\bf Figure 5.} 
The angular distribution  for $t=0.3$ GeV$^2$ 
integrated over $Q^2>1$ GeV$^2$ and $W>2$ GeV.}
\label{phi_dist}
\end{figure}

\begin{figure}[b!]
\begin{center}
\begin{tabular}{ccc}
\mbox{\epsfig{figure=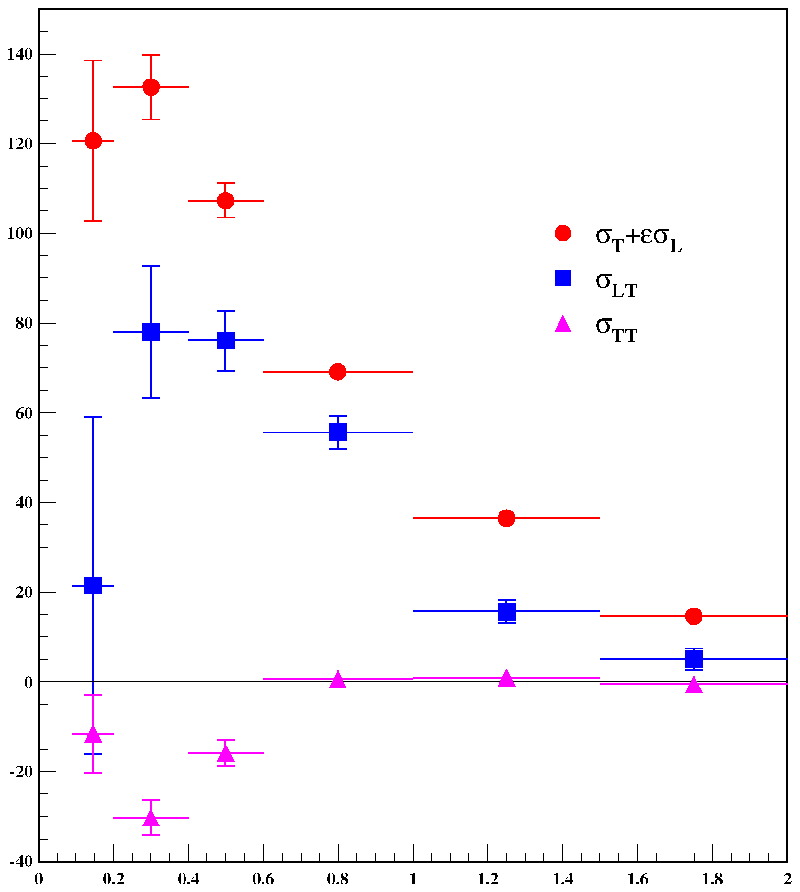,width=6cm,height=6cm}}&
\mbox{\epsfig{figure=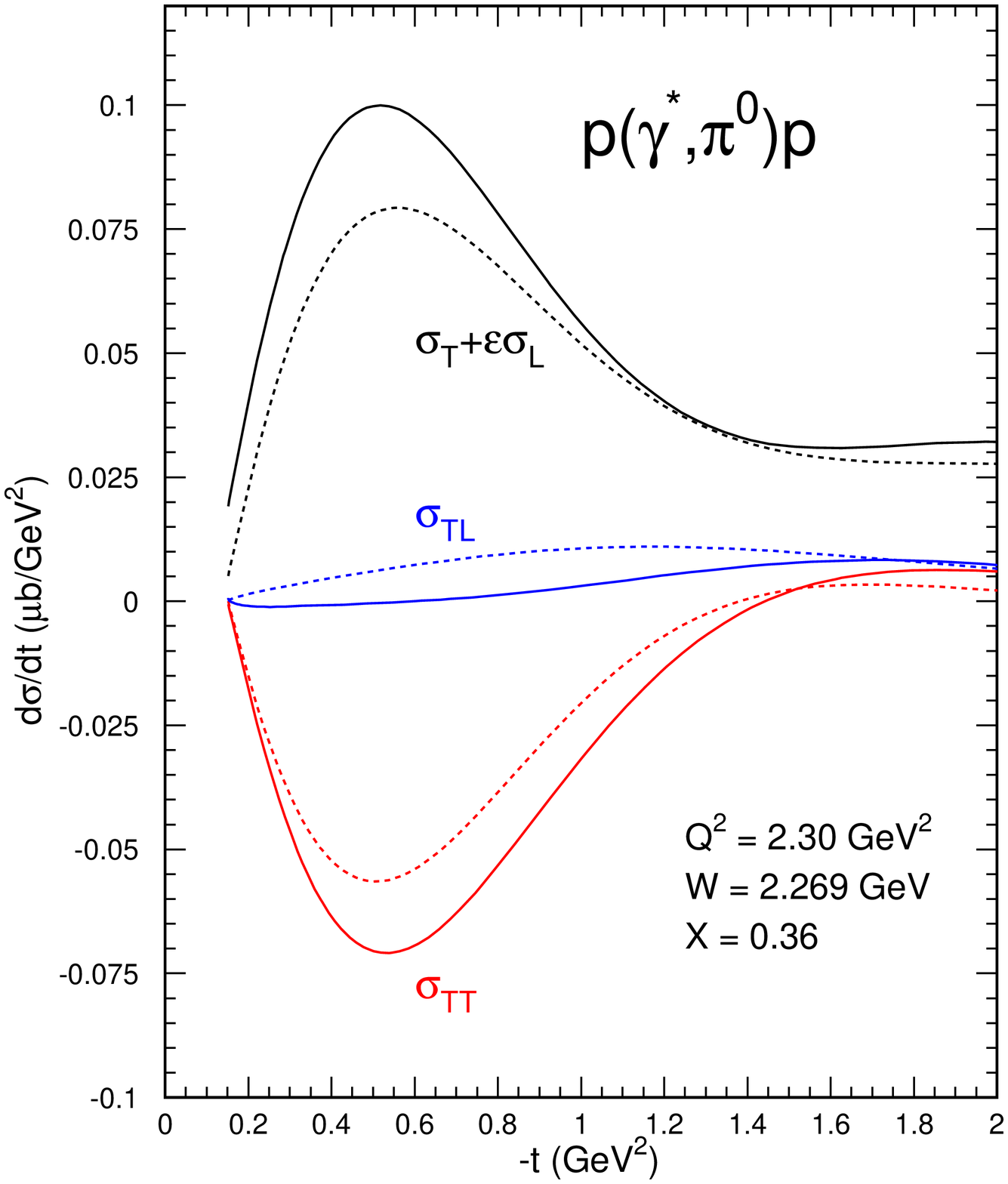,width=6cm,height=6cm}}\\
{\bf(a)}& {\bf(b)}
\end{tabular}
\end{center}
{\small{\bf Figure 6a.} 
The separated structure functions $\sigma_T+\epsilon \sigma_L $, $ \sigma_{TT}$
and $\sigma_{LT}$ as a function of $-t$ at $Q^2=2.3$ \ GeV$^2$ obtained with the CLAS
spectrometer (very preliminary, arbitrary units). 
}\\
{\small{\bf Figure 6b.} 
The results of a Regge model calculation~\cite{regge}.}
\label{t_dist}
\end{figure}
\begin{figure}[b!]
\begin{center}
\begin{tabular}{ccc}
\mbox{\epsfig{figure=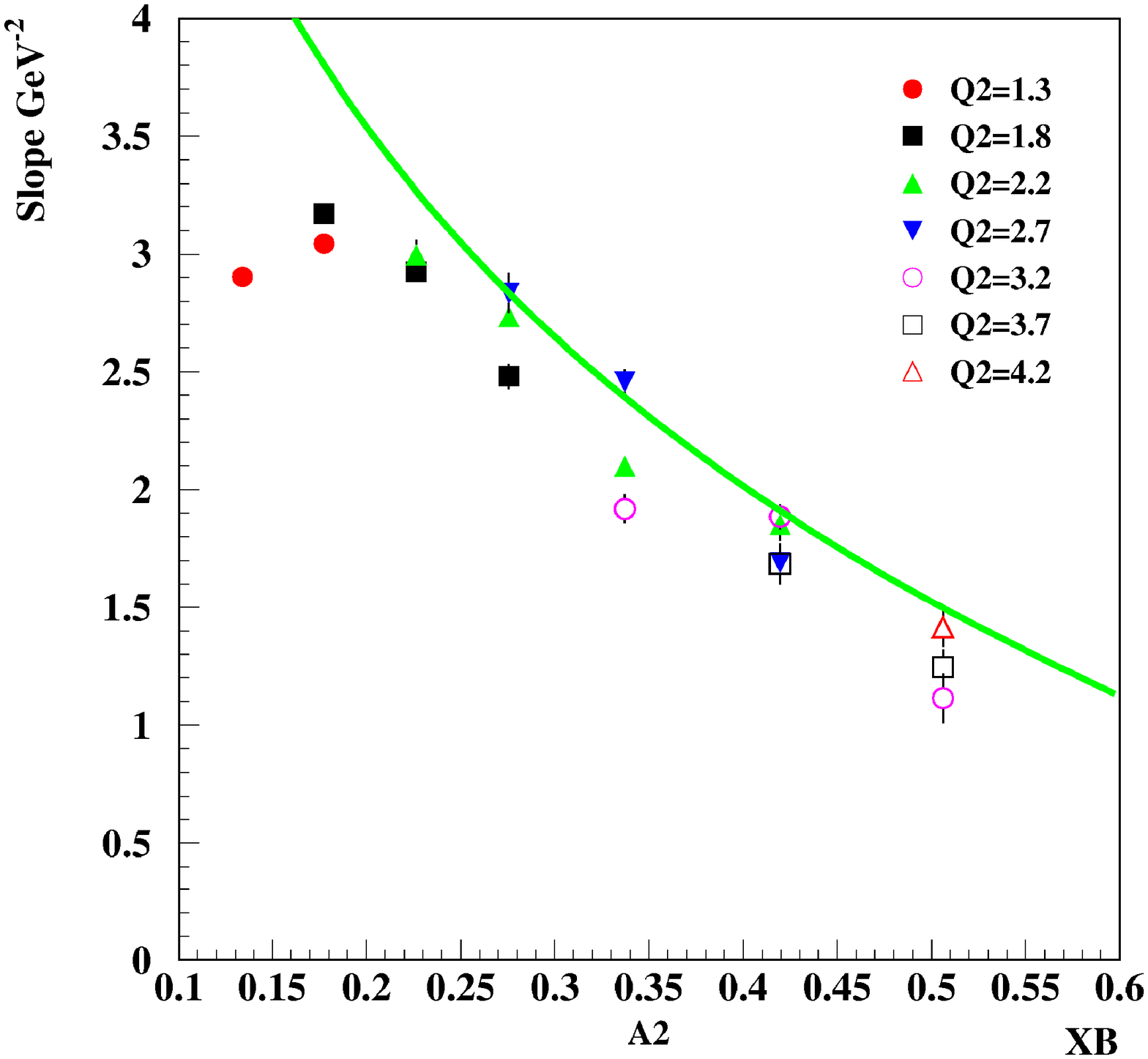,width=6cm,height=6cm}}\\
\end{tabular}
\end{center}
{\small{\bf Figure 7.} 
The experimental slope parameters $B$\ (very preliminary) obtained from fits to the data for 
various values of $Q^2$ and 
$x_B$ with the function $d\sigma/dt \propto exp\left(B(x_B,Q^2)t\right)$. 
The solid curve is
the  Regge inspired parametrization $B(x_B)=2\alpha ln(1/x_B)$\ with  $\alpha = 1.1$.}
\label{tslope}
\end{figure}

The separated structure functions versus  $t$ for   $Q^2$ = 2.3 GeV$^2$  are
shown in 
Fig.~6.
The cross sections are in arbitrary units and  radiative corrections have not been applied.
  It is observed that all the structure functions have significant
non-zero values.  $\sigma_{LT}$ is comparable to   $\sigma_T + \epsilon \sigma_{L}$
which implies  that there are  significant contributions of transverse amplitudes
at these relatively low values of $Q^2$,
so the  factorization cannot
be applied. However, one may analyze these data in terms of a hadron based models
such as Regge phenomenology~\cite{regge}.
Fig.~6  shows the results of such a calculation, which qualitatively 
follows the sign of the separated structure functions, but not always the shape.

The Fourier transformation of the GPDs gives information about the impact 
parameter $\vec{b}_\perp$ dependence
of  parton distributions. The Fourier transformations are given by

\begin{equation}
F(x,\vec b_\perp)\propto \int {\frac{d^2\vec \Delta_\perp}{ 2\pi}} e^ {i\vec \Delta_\perp \vec b_\perp} \tilde H(x,0,\Delta_\perp^2)
\label{eq:F}
\end{equation}

 Due to the significant contribution of transverse amplitudes at the current 
kinematics we do not 
 have access to GPDs. However, we can apply a Fourier transformation to 
the cross sections to get impact parameter information.
Slope parameters $B$ have been extracted by fitting the $t$ distributions
using the  parametrization $d\sigma/dt \propto exp\left(B(x_B,Q^2)t\right)$. The result is shown in 
Fig.~7. 
Note that  $B$ does not appear to significantly
depend on  $Q^2$.

In a Regge inspired GPD model, the $x_B$ dependence of the slope parameter is
given by  $B(x_B)=2\alpha ln(1/x_B)$, with $\alpha \sim 1$ \cite{radyushkin}. The curve  in 
Fig.~7
is a plot of this parametrization for $B$.  Remarkably, this curve appears to
accurately account for the data with no further parameters or normalization applied.


 For the interpretation in terms of the impact parameter, 
the $\Delta_\perp^2$ slope is relevant,
where $\Delta_\perp^2$ is the transverse component of the momentum transfer
($\Delta^2=t$),
and the slope parameter is $B_\perp={\frac{B}{ {1-x_b}}}$ \cite{burkardt}. 
The fact that the t-slope goes to zero for large $x_B$ may be purely
kinematical. 
However, even taking into account this factor, we note that
$B_\perp$ falls with $x_B$ in the region $x_B$ from 0.1 to 0.5 where we have experimental data.  
This implies that 
the impact parameter distribution is broadest at lowest $x_B$
and becomes  narrower at increasing $x_B$.

\begin{figure}[ht]
\begin{center}
\begin{tabular}{ccc}
\mbox{\epsfig{figure=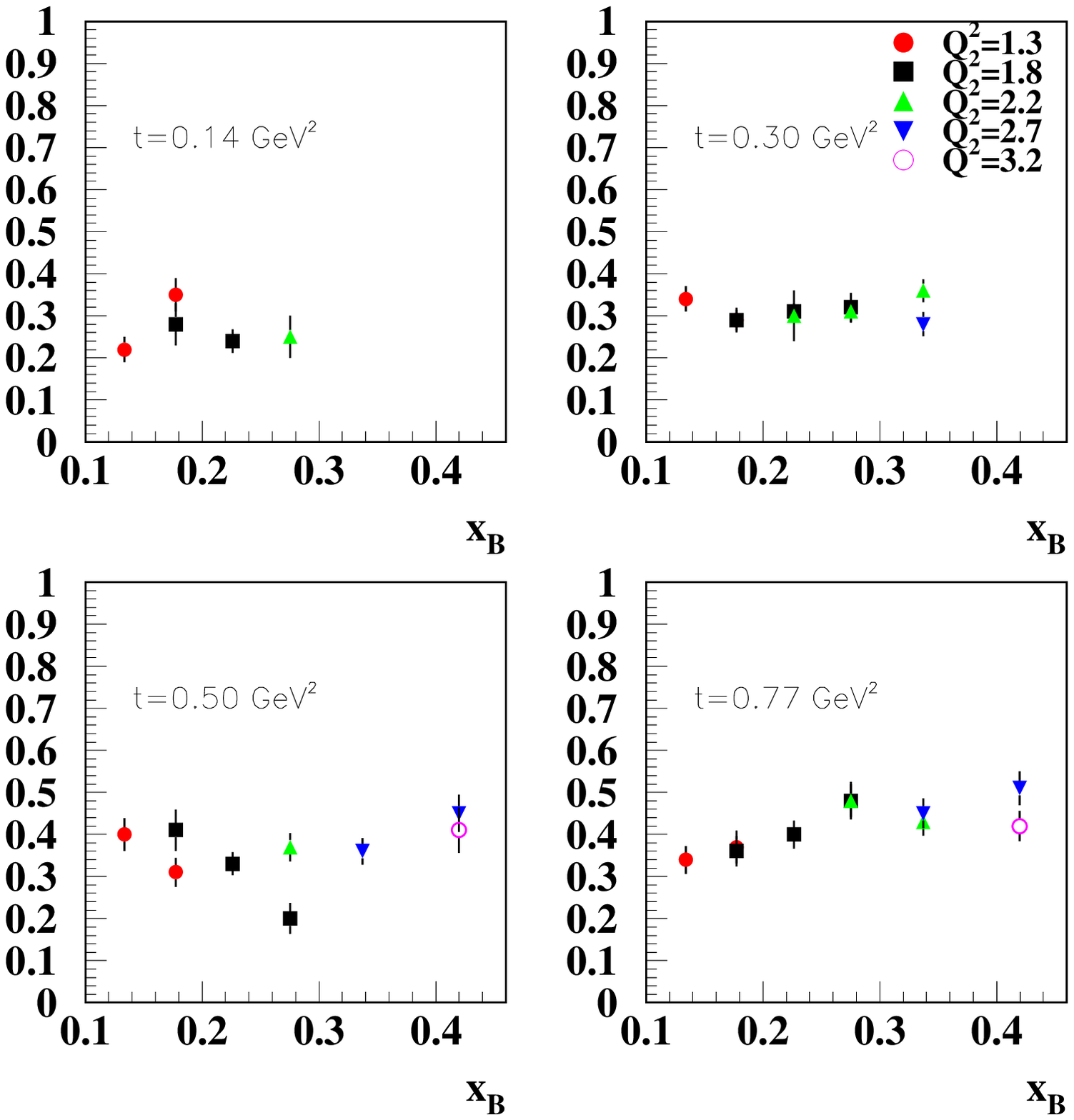,width=10cm,height=10cm}}\\
\end{tabular}
\end{center}
{\small{\bf Figure 8.} 
$\eta$ to $\pi^0$ cross sections ratio as a function of $x_B$ for different values of 
$Q^2$ and $t$ (very preliminary).}
\label{rat}
\end{figure}


%
%
\begin{figure}[ht]
\begin{center}
\begin{tabular}{ccc}
\mbox{\epsfig{figure=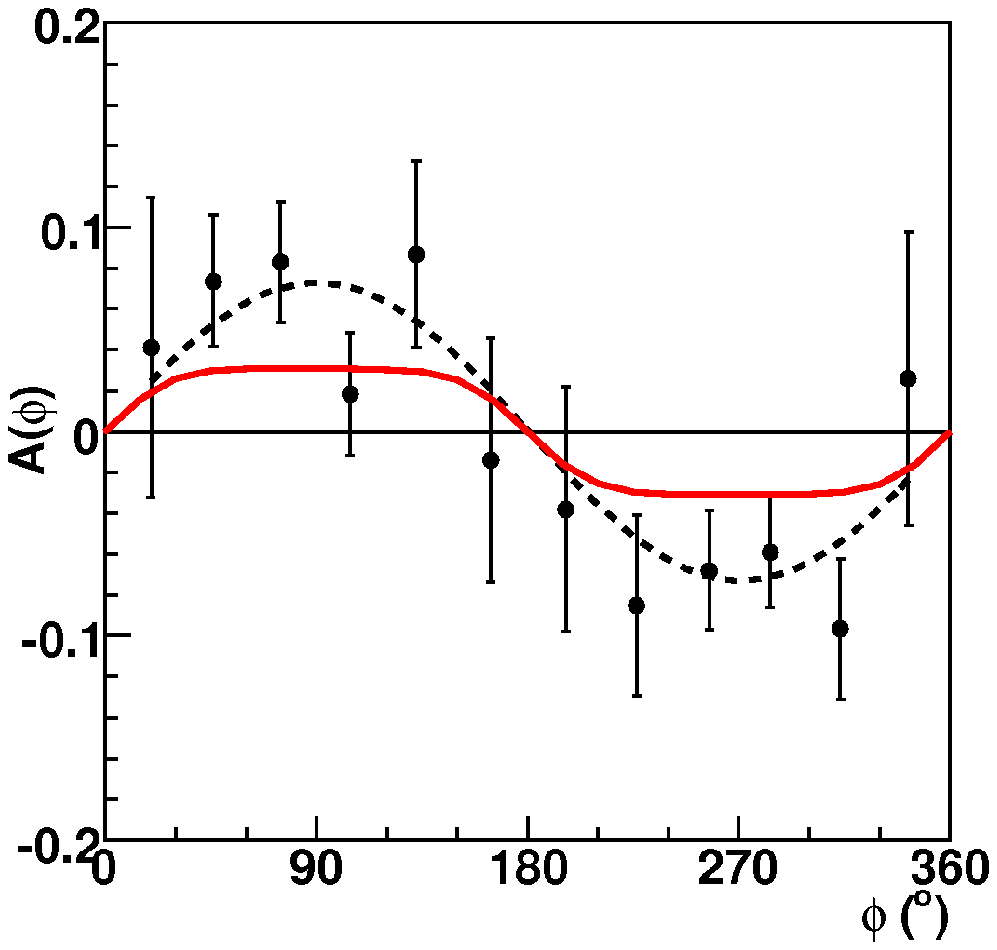,width=6cm,height=6cm}}\\
\end{tabular}
\end{center}
{\small{\bf Figure 9.} 
The angular distribution of the BSA  for $\pi^0$ at $Q^2=2 $\ GeV$^2$, 
$t=-0.3$ GeV$^2$, and $x_B=0.25$. The dashed curve is a fit to the function $A=\alpha \sin\phi$
and the solid curve is the result of a Regge model~\cite{regge} calculation. }
\label{aphidist}
\end{figure}

\subsection*{The ratio of cross sections for $\eta$ and $\pi^0$.}
As it was noted in the introduction the ratio of cross sections may play an 
important role due to the {\it precocious factorization}.
This ratio is presented in Fig.~8 for the different values of
$t$ and $Q^2$ as a function of $x_B$. Note that this ratio is almost independent of $x_B$
and varies from 0.3 to 0.4 with increasing $t$. This is in contrast with the prediction
\cite{precocious} (see Fig.~2), where this ratio is equal to 1.
However, we can not compare directly with   \cite{precocious}
since  $\sigma_L$ and $\sigma_T$ were not separated.

\subsection*{Beam spin asymmetry.}
The beam spin asymmetry (BSA) is defined by 
\begin{equation}
 A={\frac{\stackrel{\rightarrow}{\sigma}  -  \stackrel{\leftarrow}{\sigma}  }{\stackrel{\rightarrow}{\sigma}   + \stackrel{\leftarrow}{\sigma} } }
~\sim \alpha sin\phi . 
\label{asym}
\end{equation}
\noindent From Eq.~\ref{eq:sigma} the beam spin asymmetry directly yields the $L-T$ 
interference structure function $\sigma^\prime_{LT}$. Any observation of a non-zero BSA
 would be indicative of an L-T interference.
If  $\sigma_L$ dominates, then $\sigma_{LT}, \sigma_{TT}$, and $\sigma^\prime _{LT}$
should be small.  An example of a $\phi$ distribution of the BSA for $\pi^0$ and $\eta$
production at a particular kinematic bin is shown in 
Fig.~9.


Sizable beam-spin asymmetries for exclusive $\pi^0$ and $\eta$ mesons 
electroproduction have been measured above the resonance region for the first time.
These non-zero asymmetries imply that both transverse and longitudinal 
amplitudes participate in the process. However, the results of a  Regge 
model calculation qualitatively describe the 
experimental data too.  

\subsection*{Conclusion.}
Cross sections and asymmetries for the $\pi^0$  and $\eta$  exclusive electroproduction 
in a very wide kinematic range of $Q^2$, $t$\ and $x_B$ have been measured and initial 
analyses already are showing remarkable results.
These data will help us to better understand  the transition from soft to hard mechanisms.
Initial results  show that both transverse and longitudinal amplitudes participate in 
the exclusive processes at currently accessible kinematics. 
The $\pi^0/\eta$ cross section ratio will check the hypothesis of precocious scaling.

We view the work  presented here as leading into the 
program of the Jefferson Lab 12 GeV upgrade. The increased energy and luminosity
will allow us to  make the analysis presented here at much higher $Q^2$ and $x_B$ as well as
to perform Rosenbluth $L/T$ separations.
In parallel, we pose the following theoretical questions.
What does the $t$-slope $B(Q^2,x_B)$ tell us? What can we learn from the $Q^2$
evolution of the cross sections? Can the measurement of
$\sigma_L$, $\sigma_T$,$\sigma_{LT}$,  $\sigma_{TT}$ and $R\equiv \sigma_L /\sigma_T$ \ 
constrain GPDs within the approximations and 
corrections which have to be  made due to non-asymptotic kinematics? 
How big are the corrections?

We acknowledge the outstanding efforts of the staff of the
Accelerator and Physics Divisions at Jefferson Lab
that made this experiment possible.
We also acknowledge useful discussions with 
H.~Avakian, 
M.~Burkardt, 
V.~Burkert,
R.~De~Masi,
M.~Garcon,
F-X~Girod, 
G.~Goldstein, 
L.~Elouadrhiri,
J-M.~Laget,
S.~Liuti,
S.~Niccolai,
R.~Niyazov,
S.~Stepanyan,
M.~Strikman,
A.~Radyushkin,
I.~Strakovski,
C.~Weiss,
and
B.~Zhao.
This work was supported by
the U.S. Department of Energy
and National Science Foundation.
The Jefferson Science Associates (JSA) operates the
Thomas Jefferson National Accelerator Facility for the United States
Department of Energy under contract DE-AC05-06OR23177.

\end{document}